\documentclass[twocolumn,aps,prl,showpacs,amsmath]{revtex4}
\usepackage{graphicx}
\usepackage{dcolumn}

\begin{document}
\title{A surprising relation between double exchange and Heisenberg model
spectra: Application to half-doped manganites}

\vspace{2cm}

\author{Roland Bastardis, Nathalie Guih\'ery, Nicolas Suaud}
\affiliation{Laboratoire de Physique Quantique, IRSAMC/UMR5626,
Universit\'e Paul Sabatier, 118 route de Narbonne, F-31062 Toulouse
Cedex 4, FRANCE} 
\date{\today}

\begin{abstract}

The Zener polarons recently found in half-doped manganites are usually seen as
mixed valence entities ruled by a double exchange Hamiltonian involving only
correlated electrons of the metals. They can however be considered as
ferrimagnetic local units if the holes are localized on the bridging oxygen
atoms as implicitely suggested by recent mean-field {\it ab initio} calculations. 
In the latter case, the physics is ruled by a Heisenberg
Hamiltonian involving magnetic oxygen bridges. This paper shows that the spectra
resulting from the resolution of both models are {\it analytically identical}.
This single resulting model spectrum accurately reproduces the spectrum of 
Zener polarons in Pr$_{0.6}$Ca$_{0.4}$MnO$_3$ manganite studied by means of
explicitely correlated {\it ab initio} calculations. Since the physics supported
by each model are different, the analysis of the exact Hamiltonian ground state
wave function should {\it a priori} enables one to determine the most appropriate
model. It will be shown that neither the spectrum nor the wavefunction analysis
bring any decisive arguments to settle the question. Such undecidability would
probably be encountered in experimental information. 

\end{abstract}
\maketitle

Strongly correlated materials have attracted the interest of chemists and
physicists for their fascinating macroscopic properties\cite{bed,jonker,isobe}.
Adequate treatments of collective effects as well as the resolution of the
microscopic origin of these properties are crucial to the understanding of these
materials. Both alike require the determination of accurate effective
Hamiltonians. In doped manganites as well as in many oxydes, the various phase
transitions occuring under doping are not fully understood. As well, relevant
and quantitative effective Hamiltonians able to reproduce the microscopic order
in the most complex phases are rarely available.

A recent crystal determination of Pr$_{0.6}$Ca$_{0.4}$MnO$_3$
manganite\cite{daoud} has revealed the presence of Zener polarons {\it i.e.}
trappings of holes (or electrons) within pairs of Manganese sites. The Zener
polaron was initially assumed to be ruled by a double exchange
model\cite{daoud,coen} giving rise to an octet ground state of the dimer. In
this description each hole is delocalized over two Mn sites resulting in a
resonance between Mn$^{3+}$O$^{2-}$Mn$^{4+}$ and Mn$^{4+}$O$^{2-}$Mn$^{3+}$.
However, recent {\it ab initio} calculations on the closely related
La$_{0.5}$Ca$_{0.5}$MnO$_3$\cite{var3,var4} show {\it an} important O to Mn charge
transfer, resulting in a localization of the holes on the bridging oxygens. 
This charge distribution suggests a dominant purely magnetic local order
(Mn$^{3+}$O$^{-}$Mn$^{3+}$) in which the Zener polarons would be ferrimagnetic
entities involving an antiferromagnetic coupling between the magnetic oxygen and
the high spin manganese centers. In the latter case the model Hamiltonian which
provides a relevant description of the local electronic order is a Heisenberg
Hamiltonian. The holes localization has been a matter of debate in the
literature and concerns a large number of materials like doped cuprates
\cite{zhang}(where there is at most one unpaired electron per metal resulting in
a $t$-$J$ model), nickelates or manganites (in which there are several
open-shells per center leading to a double exchange interaction). Besides, as
far as the correlated electrons are only localized on the metallic centers, the
comparison of the models has been the subject of intense discussion and
controversy which ended up with the demonstration that the two models were
strictly different and lead to different spectra\cite{gir2}. The aim of this
paper is to show that when considering a magnetic oxygen in the Heisenberg model
the analytical resolution of the two Hamiltonians leads to identical spectra.
For this purpose we have derived an analytical solution of the Heisenberg
Hamiltonian energies for three magnetic centers.

Since the discrimination between the two models is impossible from the spectrum
determination, we will then adress this question from wavefunction analysis. 
Indeed, the projection of the {\it ab initio} ground state wavefunction onto
both model spaces should {\it a priori} determine the choice. However, it will
be shown that when the charge of the oxygen is around -1.5, the wavefunction
analysis does not infer any decidable argument. In order to illustrate this
assertion, we have extracted the two corresponding models from the correlated
{\it ab initio} calculated spectrum of the Pr$_{0.6}$Ca$_{0.4}$MnO$_3$ manganite
Zener polaron and analyzed both its spectrum and its ground state wave
function.

The here-developped solutions are analytical and could be applied to any
material having a bridging ligand and an odd number of correlated electrons per
metallic dimer. Thus, the here-developed equations have been generalized to any
number n of open-shells per center. The first considered model, is the Zener
double exchange model\cite{zener} including the Heisenberg-type
antiferromagnetic contribution proposed by Papaefthymiou and
Girerd\cite{gir1,gir2}. It is likely to describe configurations compatible with
a M(d$^{(n-1/2)}$)O$^{2-}$M(d$^{n-1/2}$) mixed valence electronic structure
resulting from a resonance between M(d$^{(n)}$)O$^{2-}$M(d$^{n-1}$) and
M(d$^{(n-1)}$)O$^{2-}$M(d$^{n}$) where $n$ and $n$-1 are the numbers of open
shells on the two metallic centers. The model is based on the idea that the
spectrum can be reproduced by considering the metallic ions in their atomic
ground states and two effective interactions, namely the hopping integral $t$ of
the extra electron between the two metal-centered orbitals and an overall
Heisenberg exchange integral $J$ between the electrons of the other open shells.
The delocalization of the extra electron favors the highest spin state (of total
spin $S^Z_{max}$) while the exchange integral stabilizes the low and
intermediate spin states for an overall antiferromagnetic contribution ($J<0$).
In the particular case of a symmetric homonuclear bimetallic complex, the
eigenenergies of the double exchange model are analytically known. This
Hamiltonian generates two eigenstates for each value of the total spin $S$, the
energies of which $E^Z(S,\pm)$ are given by the expression:

\begin{figure}[t]
\centerline{\rotatebox{0}{\resizebox{6cm}{6cm}{\includegraphics{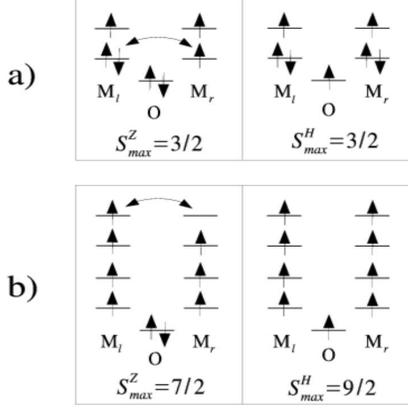}}}}
\caption{Sample cases of electronic distributions (left: double exchange model;
right: Heisenberg model) illustrating the two kinds of systems. In a)
$n_u^H=n_u^Z$ (Nickelates) while in b) $n_u^H=n_u^Z+2$ (Manganites).}
\label{fig-nbetats}
\end{figure}

\begin{eqnarray}
\label{enerZ}	  
E^Z(S, \pm )&=&\pm~\frac{t}{S_{max}^Z +1/2} \left(S+\frac{1}{2}\right)\\
   && -~\frac{J}{2} [S(S +1) - S_{max}^Z(S_{max}^Z+1)]\nonumber
\end{eqnarray}
where the zero of energy is taken as the mean value of the highest spin states
energies.

If the electronic structure is dominated by M(d$^n$)O$^{-}$M(d$^n$) charge
tranfer configurations, a Heisenberg Hamiltonian is expected to rule the physics
of the system. Its specific form is ${\bf H}^{H}=-J_1{\bf S}_{l}.{\bf
S}_O-J_1{\bf S}_{r}.{\bf S}_O-J_2{\bf S}_{l}.{\bf S}_{r}$ where $J_2$ is the
overall spin exchange interaction between the left ({\bf S}$_l$) and right ({\bf
S}$_r$) metal spin, and $J_1$ the magnetic interaction between the metal and the
oxygen ions. The Heisenberg model space is also built from products of atomic
ground states. The ${\bf S}^2$ operator can be written as 
\begin{eqnarray}
{\bf S}^2 &=&{\bf (S}_l+{\bf S}_r+{\bf S}_O)^2 \\
 &=& {\bf S}^2_r+{\bf S}^2_l+{\bf S}^2_O
 +2({\bf S}_l.{\bf S}_O +{\bf S}_r.{\bf S}_O +{\bf S}_r.{\bf S}_l)\nonumber
\end{eqnarray}

Replacing ${\bf S}_r.{\bf S}_l$ by its expression as function of ${\bf S}^2$ 
in the Hamiltonian, one gets 
\begin{eqnarray}
{\bf H}^{H}= -(J_1-J_2) {\bf S'.S}_O -\frac{J_2}{2}({\bf S}^2-{\bf S}_r^2-{\bf
S}_l^2-{\bf S}_O^2)
\end{eqnarray}
where ${\bf S'=S}_r+{\bf S}_l$. As ${\bf S'.S}_O=\frac{1}{2}({\bf
S}^2-{\bf S'}^2-{\bf S}_O^2)$, the Hamiltonian can be rewritten as:
\begin{eqnarray}
\hspace*{-4mm}{\bf H}^{H}=-\frac{(J_1-J_2)}{2}[{\bf S}^2-{\bf S'}^2-{\bf
S}_O^2]-\frac{J_2}{2}[{\bf S}^2-{\bf S}_l^2-{\bf S}_r^2-{\bf S}_O^2]
\end{eqnarray} 

Since the oxygen has only one unpaired electron, {\it i.e.} $S_O=1/2$, the
values of $S'$ can only be $S'=S \pm \frac{1}{2}$ and an analytical expression
of the eigenenergies can be found:
\begin{eqnarray}
\label{enerH}
E^{H}(S,\pm)& = & \frac{J_1-J_2}{2}\left[1\pm \left(S+\frac{1}{2}\right)\right] \\
& & -\frac{J_2}{2}[S(S+1)
-\frac{1}{2}(S_{max}^H+\frac{1}{2})^2-\frac{1}{4}]\nonumber
\end{eqnarray} 
where $S_{max}^H=S_l+S_r+\frac{1}{2}$. For the highest spin multiplicity state,
only the $E^{H}(S,-)$ root has a physical meaning since $S'$ has the single
value $S'=S-·\frac{1}{2}$.

In order to compare the two models, the same zero of energy must be used in
equations (\ref{enerZ}) and (\ref{enerH}). Let us use the double exchange model
zero of energy. Its expression as a function of the Heisenberg model is:
\begin{eqnarray}
\hspace*{-2mm}E_0 &=& \frac{E^H(S_{max}^Z,+)+E^H(S_{max}^Z,-)}{2}\\
\hspace*{-2mm} & = & \frac{J_1-J_2}{2}-
\frac{J_2}{2}
\left[S_{max}^Z(S_{max}^Z+1)-\frac{1}{2}\left(S_{max}^H+\frac{1}{2}\right)^2-\frac{1}{4}\right] \nonumber
\end{eqnarray}

The relation between $S^Z_{max}$ and $S^H_{max}$ is fixed by the number $n_{u}$
of unpaired electrons in the two models ($n_u^H$ and $n_u^Z$ for the Heisenberg
and double exchange models, respectively). If $n_{u}^H=n_u^Z$, then
$S^H_{max}=S^Z_{max}$. On the contrary, if $n_u^H=n_u^Z$+2 then
$S^H_{max}=S^Z_{max}+1$. These two situations are illustrated in
figure~\ref{fig-nbetats}. The first one (schemes a) corresponds, for instance, to
dopped nickelates. The delocalized electron of the double exchange
model (left scheme) jumps from a doubly occupied orbital towards a singly occupied one. 
The second situation (schemes b) is representative of half-doped manganites. The hopping electron of the
double exchange model jumps from a singly occupied orbital towards the 
empty orbital of the other manganese.

The Heisenberg energies shifted by the mean value $E_0$ of the $S_{max}^Z$
states become:
\begin{eqnarray}
\label{enerH2}
E^H(S,\pm)&=&\pm~\frac{J_1-J_2}{2}\left(S+\frac{1}{2}\right)\\
   &&-~\frac{J_2}{2}\left[S(S+1) -S_{max}^Z(S_{max}^Z+1)\right]\nonumber
\end{eqnarray}

Thus, from comparisons between equations~(\ref{enerZ}) and~(\ref{enerH2}), an
important conclusion is that the energies of the two models are exactly the same
if one replaces $t$ and $J$ by

\begin{eqnarray}
t = \frac{1}{2}(J_1-J_2)\left(S_m^Z+\frac{1}{2}\right) \hspace{5mm} \mathrm{and}
\hspace{5mm}J = J_2
\end{eqnarray}

Let us notice that when $S_{max}^H=S_{max}^Z+1$, the energy of the $S_{max}^H$
state cannot be obtained in the double exchange model, since this state does not
belong to the model space. Its energy is only given by the Heisenberg
expression. That means, in dopped manganites case, that the double exchange
model contains two doublet, two quartet, two sextet and two octet ($S=7/2$)
states whereas the Heisenberg model also describes a decuplet state ($S=9/2$).
On the contrary, when $S_{max}^H=S_{max}^Z$, the $E^H(S_{max}^H,+)$ solution is
meaningless for the Heisenberg model (see eq.~\ref{enerH} and its explanation)
whereas $E^Z(S_{max}^Z,+)$ is a solution of the double exchange model.
Nevertheless, we still have $E^H(S_{max}^H,+)=E^Z(S_{max}^Z,+)$.  So, the
complete description of the spectrum is possible from the Heisenberg expression
while the Heisenberg model space does not contain the $E^H(S_{max}^H,+)$ state.
One therefore has the unexpected situation in which two model Hamiltonians
working on different model spaces generate spectra which coincide except for one
state. If $S_{max}^H=S_{max}^Z+1$ the double exchange spectrum is included in
the Heisenberg one while it is the Heisenberg spectrum which is fully described
by the double exchange model when $S_{max}^H=S_{max}^Z$. 

\begin{table}[b]
\begin{tabular}{cccccc}
   & $J_1$ & $J_2$ & $t$ & $J$ & error (\%)\\
Heisenberg & 0.59 & 0.07  & - & - & 3.3   \\
Double exchange & - & - & 1.05  & 0.07  & 3.3 \\
\end{tabular}
\caption{Effective interactions (in eV) of the two models extracted from the
{\it ab initio} spectrum and errors of the corresponding predicted spectrum
compared to the {\it ab initio} one.}
\label{table1}
\end{table}

\begin{figure}[h]
\centerline{\rotatebox{0}{\resizebox{8cm}{7cm}{\includegraphics{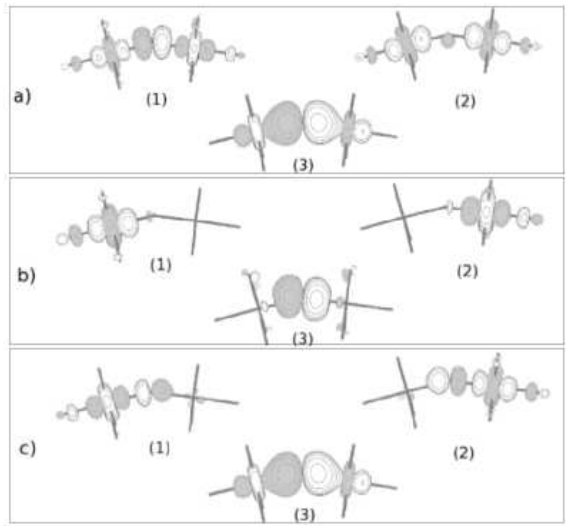}}}}
\caption{Canonical delocalized orbitals~(a); strongly localized
orbitals~(b) and left or right semi-localized orbitals~(c).}
\label{figOM}
\end{figure}

In order to check the relevance of the two considered models on a real material,
let us turn now to the explicitely correlated {\it ab initio} study of the
Pr$_{0.6}$Ca$_{0.4}$MnO$_3$ manganite. Since the choice between the two models
turns out to be impossible from the only determination of the spectrum, the
analysis of the ground state wavefunction becomes compulsory in order to
discriminate between them. The Mn$_2$O$_{11}^{15-}$ cluster (constituted of the
dimer of Mn sites surrounded by the oxygen atoms of their coordination sphere)
is analyzed using the embedded cluster method. In order to reproduce the crystal
environment, the cluster is embedded in a bath that reproduces its main effects
(Madelung and Pauli exclusion)\cite{coen,rolandd}. Then the cluster is studied
using correlated \textit{ab initio} calculations and large basis sets
\cite{ano}. MS-CASPT2 calculations~\cite{caspt21} (the many-body second order
perturbation scheme of the MOLCAS code\cite{molcas} which has been successfully
applied to many spectroscopic studies\cite{roos}) are performed (for a more
detailed description, see references \cite{coen} and \cite{rolandd}). The
so-obtained spectrum constituted of height states compares well with the model
spectrum\cite{rolandd}. Table~\ref{table1} reports the optimized parameters and
the mean error per states of the two models. This error is calculated as the
energy difference between the computed (\textit{ab initio}) states and those
reproduced by the model divided by the spectrum width and the total number of
states. These results show that both models appear to be appropriate to the
description of the polaron physics (it is interesting to note that a dramatic
improvment of the model resulting in a 0.7 \% of error is obtained when the
Non-Hund states are explicitely considered, as shown in ref.~\cite{rolandd}).
The large value of $J_1$ reveals a partially covalent interaction between the Mn
atoms and the bridging oxygen. However, the energetic position of the decuplet (
2$S$+1=10) state in the narrow vicinity of the highest octet state is in perfect
agreement with the prediction of the Heisenberg model\cite{rolandd}. This result
attests of the partially magnetic character of the oxygen. This intermediate
character of the metal-oxygen interaction is confirmed by the shape of the
symmetry-adapted natural orbitals which are represented in figure~\ref{figOM}a.
The delocalization between the metals and the oxygen ligand is large in both the
essentially metallic orbital (1 and 2) and the essentially 2p$_\sigma$ oxygen
centered orbital (3). The occupation number (1.6) of the essentially 2p$_\sigma$
oxygen centered orbital (3) is in between the two limit values of 2.0 in the DE
model and 1.0 in the Heisenberg one.

In order to discriminate between the two hamiltonians we have compared the
weight of the ground state model wavefunctions in the exact Hamiltonian ground
state wavefunction. Model Hamiltonians assume localized orbitals sets which must
be properly defined. The symmetry-adapted orbitals set is not appropriate to the
analysis of the Heisenberg wave functions due to the large delocalization of
orbitals (1) and (3) (see fig.~\ref{figOM}a). In order to calculate the weight
of the $ \rm Mn^{3+}O^-Mn^{3+}$ electronic configuration, orbitals (1), (2) and
(3) have been relocalized using Boys localizations\cite{Boys}. The three
strongly relocalized orbitals are represented in Figure 2b.

The Heisenberg Octet ground state wave function is : 
\begin{eqnarray*}
\Psi ^H\left(\frac{7}{2},-\right)=
&\frac{\sqrt{2}}{6}[&\mathrm{(Q_1)}_l\mathrm{D_{+1/2}(Q_2)}_r -4
\mathrm{(Q_2)}_l\mathrm{D_{-1/2}(Q_2)}_r\\
&& +\mathrm{(Q_2)}_l\mathrm{D_{+1/2}(Q_1)}_r]
\end{eqnarray*}

\noindent where $\mathrm{(Q_2)}_l\mathrm{D_{-1/2}(Q_2)}_r$ is the product of the
local left and right $Sz=2$ quintet components $ \rm (Q_2)$ on the two Mn sites
and a local $Sz=-1/2$ doublet component $\rm D_{-1/2}$ on the oxygen.
$\mathrm{(Q_1)}_r\mathrm{D_{+1/2}(Q_2)}_l$ (resp.
$\mathrm{(Q_2)}_r\mathrm{D_{+1/2}(Q_1)}_l$) is the product of the right
$\mathrm{(Q_1)}_r$ (resp. left $\mathrm{(Q_1)}_l$) quintet $Sz=1$ component on
Mn$_r$ (resp. Mn$_l$) and $\rm D_{+1/2}$ is the $Sz=+1/2$ doublet component on
O. 

The projection of the calculated ground state wave function onto the Heisenberg
ground state gives a weight of 0.74 showing the adequation of this model
Hamiltonian to reproduce the polaron physics.

The same wave function analysis has been performed on the double exchange model
space. The double exchange octet ground state wave function is :
\begin{eqnarray*}
\Psi ^Z\left(\frac{7}{2},-\right) & = &
\frac{1}{\sqrt{2}}[\mathrm{(Q_2)}_l\mathrm{(\mathcal{Q}_{3/2})}_r
+\mathrm{(\mathcal{Q}_{3/2})}_l\mathrm{(Q_1)}_r]
\end{eqnarray*}
\noindent where $\mathrm{\mathcal{Q}_{3/2}} $ is the $S_z=3/2$ component of a
local quartet. This model assumes a two-electron occupied orbital centered on
the oxygen which is actually well described within the symmetry-adapted 
delocalized picture. The double exchange eigenvectors are linear combinations of
two resonant forms where the extra electron is either on a left or on a right
localized orbital. The adequate metallic orbitals are semi localized on  the
left and on the right metals with correct delocalization tails of the bridging
oxygen. The so-obtained three orbitals are represented in Figure 2c. The weight
of the calculated wave function on the double exchange model space in this
orbitals set is 0.72 (to be compared to 0.74 for the Heisenberg model),
revealing that {\it both models are equivalently suitable for describing the
physics of the polaron}. 

As a conclusion, since the spectra of the double exchange and the Heisenberg
models are identical, the only criterium to choose the most appropriate model
for a specific system should be the wavefunction projection onto both model
spaces. In limit cases where the oxygen charge is close to -1 (or -2), i.e. for
dominant $\rm M(d^n)O^{-}M(d^n)$ (or $\rm M(d^{n-1})O^{2-}M(d^n)$) electronic
structure, a wavefunction analysis would enable one to decide which model is the
most appropriated. On the contrary, for intermediate occupation of the oxygen
orbital (which is the case in many correlated materials), the exact Hamiltonian
wave function projection amplitudes onto both model spaces are almost the same
provided that one uses proper orbitals sets. Consequently, the discrimination
between the double exchange and Heisenberg model Hamiltonians, which is a matter
of numerous discussions \cite{var2,var3,var4,coen,rolandd}, is simply not
possible, neither from the spectrum nor from the wave function analysis. It is
likely that this undecidability will be encountered in the interpretation of
experimental informations.

\end{document}